\documentclass[twocolumn,american,pre,showpacs,preprintnumbers,amsmath,amssymb]{revtex4-1}
\usepackage[latin9]{inputenc}
\usepackage{amsbsy}
\usepackage{amstext}
\usepackage{amssymb}
\usepackage{graphicx}
\usepackage{esint}

\makeatletter

\usepackage{dcolumn}
\usepackage{bm}

\usepackage{babel}

\usepackage{babel}

\makeatother

\usepackage{babel}
\begin{document}
\global\long\def\bea{ 
\begin{eqnarray}
\end{eqnarray}
 }
 \global\long\def\eea{ {eqnarray}}
 \global\long\def\bit{\begin{itemize}\end{itemize}}
 \global\long\def\eit{ {itemize}}

\global\long\def\be{ 
\begin{equation}
\end{equation}
 }
 \global\long\def\ee{{equation}}
 \global\long\def\ra{\rangle}
 \global\long\def\la{\langle}
 \global\long\def\U{\widetilde{U}}


\global\long\def\bra#1{{\langle#1|}}
 \global\long\def\ket#1{{|#1\rangle}}
 \global\long\def\bracket#1#2{{\langle#1|#2\rangle}}
 \global\long\def\inner#1#2{{\langle#1|#2\rangle}}
 \global\long\def\expect#1{{\langle#1\rangle}}
 \global\long\def\e{{\rm e}}
 \global\long\def\proj{{\hat{{\cal P}}}}
 \global\long\def\tr{{\rm Tr}}
 \global\long\def\H{{\hat{H}}}
 \global\long\def\Hdag{{\hat{H}}^{\dagger}}
 \global\long\def\Lop{{\cal L}}
 \global\long\def\Ehat{{\hat{E}}}
 \global\long\def\Edag{{\hat{E}}^{\dagger}}
 \global\long\def\Shat{\hat{S}}
 \global\long\def\Sdag{{\hat{S}}^{\dagger}}
 \global\long\def\Ahat{{\hat{A}}}
 \global\long\def\Adag{{\hat{A}}^{\dagger}}
 \global\long\def\U{{\hat{U}}}
 \global\long\def\Udag{{\hat{U}}^{\dagger}}
 \global\long\def\Zhat{{\hat{Z}}}
 \global\long\def\Phat{{\hat{P}}}
 \global\long\def\Op{{\hat{O}}}
 \global\long\def\id{{\hat{I}}}
 \global\long\def\x{{\hat{x}}}
 \global\long\def\P{{\hat{P}}}
 \global\long\def\Px{\proj_{x}}
 \global\long\def\Pr{\proj_{R}}
 \global\long\def\Pl{\proj_{L}}


\title{Quantum and classical complexity in coupled maps}

\author{Pablo D. Bergamasco}

\affiliation{Departamento de Física, CNEA, Libertador 8250, (C1429BNP) Buenos 
Aires, Argentina}
\affiliation{Departamento de Física, FCEyN, Universidad de Buenos Aires,
Argentina}

\author{Gabriel G. Carlo}

\affiliation{Departamento de Física, CNEA, CONICET, Libertador 8250, (C1429BNP) Buenos
Aires, Argentina}

\author{Alejandro M. F. Rivas}

\affiliation{Departamento de Física, CNEA, CONICET, Libertador 8250, (C1429BNP) Buenos
Aires, Argentina}

\email{pablobergamasco@cnea.gov.ar,carlo@tandar.cnea.gov.ar,rivas@tandar.cnea.gov.ar}

\selectlanguage{american}%

\date{\today}

\begin{abstract}
We study a generic and paradigmatic two degrees of freedom system consisting of two coupled 
perturbed cat maps with different types of dynamics. The Wigner separability entropy (WSE) -- 
equivalent to the operator space entanglement entropy -- and the classical separability 
entropy (CSE) are used as measures of complexity. For the case where both degrees of freedom are
hyperbolic, the maps are classically ergodic and the WSE and the CSE behave similarly, growing up to higher 
values than in the doubly elliptic case. However, when one map is elliptic and the other hyperbolic, 
the WSE reaches the same asymptotic value than that of the doubly hyperbolic case, 
but at a much slower rate. The CSE only follows the WSE for a few map steps, 
revealing that classical dynamical features are not enough to explain complexity growth.
\end{abstract}


\maketitle
\section{Introduction}

\label{sec1}

Chaotic behaviour is a classical property that implies the exponential
divergence of close initial conditions. The ability to explore the 
whole available phase space, i.e. ergodicity, is indeed the main 
ingredient for statistical thermodynamics.
On the other hand, quantum mechanics is governed by the Schr\:odinger
equation whose linearity forbids exponential divergences of close
initial conditions. Also, entanglement is a quantum characteristic
that has no classical counterpart. Consequently, as recently pointed
out in \cite{nature2017}, there is a battle
between quantum and thermodynamic laws.

We mention a few contributions to this discussion. 
From the quantum to classical correspondence point of view, a pioneering 
work \cite{Patta99} has related the classically ergodic
behaviour with quantum entanglement production. Very recently a small quantum system of three 
superconducting qubits has been considered \cite{nature2016}, showing a coincidence between regions of 
high (quantum) entanglement entropy and (classical) chaotic dynamics.
On the other hand, entropy production in regular
regions has been reported \cite{Lombardi-Matzkin}. Also, in a Toda model of 
two interacting particles the chaotic and integrable cases could hardly be distinguished 
regarding entanglement generation \cite{Casati}. Also, thermalization of quantum systems according
to their type of dynamics is a subject of fundamental interest nowadays \cite{physrep2016}. 

In order to perform an explicit comparison between quantum and classical
mechanics it is of great help to have a quantity that can be calculated in both 
realms. Wigner functions represent quantum mechanics in phase space providing with a 
very suitable analogue of Liouville distributions. Recently, in the spirit of algorithmic complexity,
the WSE \cite{Benenti-Carlo-Prosen} and the CSE \cite{prosen1} have been introduced as measures of
complexity of quantum and (discretized) classical distributions, respectively. 
In this work, we propose the notion of complexity to analyze correspondence by using these 
measures to study a two degrees of freedom system. We consider two 
coupled perturbed cat maps, where one of them can be seen as the
system and the other as the environment. The dynamics of these maps can be both hyperbolic (chaotic) (HH), 
both elliptic (regular) (EE), or mixed where one degree of freedom is hyperbolic
and the other is elliptic (HE-EH). 

We have found that for the HH case, 
the Wigner and Liouville distributions develop similar structures of increasing 
complexity, which are reflected in the WSE and the CSE hand by hand growth 
until a saturation value \cite{lak2002}. At the classical level, after an evolution 
of the order of the Ehrenfest time the entropy decreases due to discretization.
For the EE case, the quantum and classical measures do not always follow each other. 
The WSE and the CSE both reach lower values compared with the previous case. These results 
are similar to what it was found in \cite{nature2016} for chaotic and regular regions of phase space.
Finally, for the mixed HE case the quantum complexity saturates at the same values of the HH case, 
although the growth rate is much slower. The classical complexity only grows during the first few map steps 
and then decreases, unable to reach the quantum asymptotics. 
In this way, we can observe that one hyperbolic degree of freedom is enough to generate 
high values of complexity (entanglement). The classical behaviour differs and quantum mechanisms of 
complexity growth play a main role.

This paper is organized as follows: in Section \ref{sec2} we explain the concepts of WSE and CSE, and how they are 
used in our study. In Section \ref{sec3} we present our model with a brief discussion on its properties. 
In Section \ref{sec4} we explain our results in detail and, in Section \ref{sec5} we state our conclusions. 

\section{Wigner and classical separability entropies}

\label{sec2}

A state of a quantum system is described by means of the density operator
$\hat{\rho}$ acting on the Hilbert space $\mathcal{H}$ such that
$\text{Tr}\left(\hat{\rho}\right)=1$. This density operator is a
vector belonging to the space $B(\mathcal{H})$ of Hilbert-Schmidt
operators, where an inner product is defined as $\hat{A}.\hat{B}=\text{Tr}(\hat{A}^{\dagger}\hat{B})$
such that the norm $\|\hat{\rho}\|=\sqrt{\text{Tr}(\hat{\rho}^{2})}\le1$.
Decomposing the Hilbert space $\mathcal{H}$ as a tensor product $\mathcal{H}=\mathcal{H}_{1}\otimes\mathcal{H}_{2}$,
the density operator has a Schmidt decomposition, 
\begin{equation}
\hat{\rho}=\sum\sigma_{n}\hat{a}_{n}\otimes\hat{b}_{n},\label{SVDrho}
\end{equation}
with $n\in\mathbb{N}$, and where $\{\hat{a}_{n}\}$ and $\{\hat{b}_{n}\}$,
such that $\text{Tr}(\hat{a}_{m}^{\dagger}\hat{a}_{n})=\delta_{mn}$,
$\text{Tr}(\hat{b}_{m}^{\dagger}\hat{b}_{n})=\delta_{mn}$, are orthonormal
bases for $B(\mathcal{H}_{1})$ and $B(\mathcal{H}_{2})$, respectively.
The Schmidt coefficients $\sigma_{1}\ge\sigma_{2}\ge\ldots\ge0$ satisfy
$\sum_{n}\sigma_{n}^{2}=\text{Tr}(\hat{\rho}^{2})=\|\hat{\rho}\|^{2}$.
The operator space entanglement entropy~\cite{prosenOSEE} is then
defined as 
\begin{equation}
h[\hat{\rho}]=-\sum_{n}\tilde{\sigma}_{n}^{2}\ln\tilde{\sigma}_{n}^{2},
\quad\textrm{with}\:\tilde{\sigma}_{n}\equiv\frac{\sigma_{n}}{\|\hat{\rho}\|}.\label{eq:OSEE}
\end{equation}

Among the several representations of quantum mechanics, the Weyl-Wigner
representation performs a decomposition of the operators
that act on the Hilbert space $\mathcal{H}$ on the basis spanned
by $\widehat{R}_{\boldsymbol{x}}$, the set of unitary reflection operators
on points $\boldsymbol{x}\equiv(\boldsymbol{q},\boldsymbol{p})$ \cite{ozrep,opetor}
in a $2d$-dimensional compact phase space $\Omega=\Omega'\oplus\Omega''$.
These reflection operators are orthogonal in the sense that 
\begin{equation}
\text{Tr}\left[\hat{R}_{\boldsymbol{x}_{a}}\hat{R}_{\boldsymbol{x_{b}}}\right]=\left(2\pi\hbar\right)^{d}\;
\delta(\boldsymbol{x}_{b}-\boldsymbol{x}_{a}).\label{trR}
\end{equation}
Hence, any operator $\hat{A}$ acting on the Hilbert space $\mathcal{H}$
can be univocally decomposed in terms of reflection operators as follows
\begin{equation}
\hat{A}=\left(\frac{1}{2\pi\hbar}\right)^{d}\int d\boldsymbol{x}\ A_{W}(\boldsymbol{x})\ 
\hat{R}_{\boldsymbol{x}}.\label{rep}
\end{equation}
With this decomposition, the operator $\hat{A}$ is mapped on a function
$A_{W}(\boldsymbol{x})$, living in a $2d$-dimensional compact phase
space $\Omega$, the so called Weyl-Wigner symbol of the operator.
Using Eq.(\ref{trR}) it is easy to show that $A_{W}(x)$ can be obtained
by means of the trace operation 
\[
A_{W}(\boldsymbol{x})=\text{Tr}\left[\hat{R}_{\boldsymbol{x}}\ \hat{A}\right].
\]

The Wigner function is defined in terms of the Weyl-Wigner symbol
of the density operator,
\[
W(\boldsymbol{x})=(2\pi\hbar)^{-d/2}\rho(\boldsymbol{x})=(2\pi\hbar)^{-d/2}
\text{Tr}\left[\hat{R}_{\boldsymbol{x}}\hat{\rho}\right].
\]
Normalization of the density operator implies that:
\[
\int d\boldsymbol{x}W(\boldsymbol{x})=\text{Tr}\left(\hat{\rho}\right)=1,
\quad\textrm{while}\quad\int d\boldsymbol{x}W^{2}(\boldsymbol{x})=\|\hat{\rho}\|.
\]
Also, from the Schmidt decomposition of the density operator given in Eq.(\ref{SVDrho}),
we obtain the Schmidt (singular value) decomposition of the Wigner
function: 
\begin{equation}
W(\boldsymbol{x})=\sum_{n}\sigma_{n}a_{n}(\boldsymbol{x_1})b_{n}(\boldsymbol{x_2}),\label{eq:SVDWigner}
\end{equation}
where $\{a_{n}\}$ and $\{b_{n}\}$ are now orthonormal bases for
$L^{2}(\Omega_{1})$ and $L^{2}(\Omega_{2})$ (which are associated to the Hilbert space decomposition), such that:
\[
a_{n}(\boldsymbol{x_1})=\text{Tr}\left[\hat{R}_{\boldsymbol{x_1}}\hat{a}_{n}\right],
\quad\textrm{and}\quad b_{n}(\boldsymbol{x_2})=\text{Tr}\left[\hat{R}_{\boldsymbol{x_2}}\hat{b}_{n}\right].
\]
The \emph{Wigner separability entropy} is defined as \cite{Benenti-Carlo-Prosen}
\begin{equation}
h[W]=-\sum_{n}\tilde{\sigma}_{n}^{2}\ln\tilde{\sigma}_{n}^{2},\label{eq:Wignerentropy}
\end{equation}
where 
\begin{equation}
\tilde{\sigma}_{n}\equiv\frac{\sigma_{n}}{\sqrt{\int d\boldsymbol{x}W^{2}(\boldsymbol{x})}}.
\end{equation}
The coefficients $\{\tilde{\sigma}_{n}\}$ in Eq.(\ref{eq:Wignerentropy}) 
are then the same than those in Eq.(\ref{eq:OSEE}) and are the Schmidt 
coefficients of the singular value decomposition of 
$\tilde{W}\equiv W/\sqrt{\int d\boldsymbol{x}W^{2}(\boldsymbol{x})}$, such 
that $\tilde{W}$ is normalized in $L^{2}(\Omega)$: $\int d\boldsymbol{x}\tilde{W}^{2}(\boldsymbol{x})=1$.

The WSE $h[W]$ quantifies the logarithm of
the number of terms that effectively contribute to the decomposition of Eq.(\ref{eq:SVDWigner})
and therefore provides a measure of separability of the Wigner function
with respect to the chosen phase space decomposition. Comparing Eq.(\ref{eq:OSEE}) with 
Eq.(\ref{eq:Wignerentropy}), it is easy to see that
the WSE is equal to \emph{operator
space entanglement entropy}\cite{Benenti-Carlo-Prosen}, i.e. $h[W]=h[\hat{\rho}]$.

The main advantage of defining the separability entropy in phase space
by means of the Wigner function is that such quantity can be directly
translated to classical mechanics. The classical analogue of the Wigner
separability entropy is the CSE (or s-entropy) 
$h[\rho_{c}]$ defined in Ref.~\cite{prosen1},
where a classical phase space distribution $\rho_{c}(\boldsymbol{x})$ (discretized 
at the $\hbar$ scale)
is used instead of the Wigner function $W(\boldsymbol{x})$.
The CSE estimates the minimal amount of 
computational resources required to simulate the classical Liouvillian
evolution and grows linearly in time for dynamics
that cannot be efficiently simulated. Both the WSE and the CSE measure 
complexity of the distributions on the same footing. For our purposes this bridges 
the gap between quantum and classical mechanics.

It is worth mentioning that when the density operator $\hat{\rho}$ describes a pure state, 
$\hat{\rho}=|\psi\rangle\langle\psi|$,
there exists a simple relation between the WSE
and the entanglement content of the state 
$|\psi\rangle\in\mathcal{H}=\mathcal{H}_{1}\otimes\mathcal{H}_{2}$ \cite{Benenti-Carlo-Prosen}. 
In fact,
\[
h[W]=-2S(\hat{\rho}_{1})=-2S(\hat{\rho}_{2}),
\]
where $\hat{\rho}_{1}=\text{Tr}_{2}(\hat{\rho})$ and $\hat{\rho}_{2}=\text{Tr}_{1}(\hat{\rho})$
are the reduced density operators for subsystems 1 and 2, and $S$
is the von Neumann entropy. Since for a pure state $|\psi\rangle$ 
the von Neumann entropy of the reduced density matrix quantifies the entanglement
$E$ of $|\psi\rangle$~\cite{Nielsen,qcbook}, 
\begin{equation}
E(|\psi\rangle)=S(\hat{\rho}_{1})=S(\hat{\rho}_{2}),
\end{equation}
the WSE is twice the \emph{entanglement entropy}
$E(|\psi\rangle)$: 
\begin{equation}
h[W]=2\,E(|\psi\rangle).
\end{equation}

\section{Model system}

\label{sec3}

Despite their simplicity, dynamical maps capture all the essential
features of different types of complicated dynamical systems. This property 
and their relatively straightforward quantization turns them into a suitable tool 
to explore quantum to classical correspondence.
The quantization of the cat map \cite{Hannay 1980} 
(a paradigmatic linear automorphism on the torus and one of the most simple models
of chaotic dynamics) has contributed to elucidate many questions in the quantum chaos area 
\cite{Hannay 1980,Ozorio 1994,Haake,Espositi 2005}.
We here investigate the behaviour of two coupled
perturbed cat maps, a two degrees of freedom system. 
These two maps can have different types of
dynamics.

Each degree of freedom is defined on the 2-torus as \cite{Hannay 1980}
\begin{equation}
\left(\begin{array}{c}
q_{t+1}\\
p_{t+1}
\end{array}\right)=\mathcal{M}\left(\begin{array}{c}
q_{t}\\
p_{t}+\epsilon\left(q_{t}\right)
\end{array}\right)
\end{equation}
with $q$ and $p$ taken modulo $1$, and 
\[
\epsilon\left(q_{t}\right)=-\frac{K}{2\pi}\sin\left(2\pi q_{t}\right)
\]
For the ergodic case we use the hyperbolic map 
\begin{equation}
\mathcal{M}_{h}=\left(\begin{array}{cc}
2 & 1\\
3 & 2
\end{array}\right),\label{eq:Mhyper}
\end{equation}
while for regular behaviour we choose the elliptic map 
\begin{equation}
\mathcal{M}_{e}=\left(\begin{array}{cc}
0 & 1\\
-1 & 0
\end{array}\right).\label{eq:Meliptic}
\end{equation}
Quantum mechanics on the torus implies a finite Hilbert space of
dimension $N=\frac{1}{2\pi\hbar}$, where positions and momenta
are defined to have discrete values in a lattice of separation $\frac{1}{N}$
\cite{Hannay 1980}. In coordinate representation the corresponding 
propagator is given by a $N\times N$ unitary matrix  
\begin{equation}
U_{jk}=A{\exp}\left[\frac{i\pi}{N\mathcal{M}_{12}}(\mathcal{M}_{11}j^{2}-2jk
+\mathcal{M}_{22}k^{2})+F\right],\label{uqq}
\end{equation}
where 
\[
A=\left[1/\left(iN\mathcal{M}_{12}\right)\right]^{1/2},  
\]
\[
F=\left[iKN/\left(2\pi\right)\right]\cos\left(2\pi j/N\right).
\]
The states $\langle q|\mathbf{q}_{j}\rangle$ are periodic combs
of Dirac delta distributions at positions $q=j/N{\rm mod}(1)$, with $j$
integer in $[0,N-1]$.

The two degrees of freedom system is defined in a four-dimensional 
phase space having coordinates $\left(q^{1},q^{2},p^{1},p^{2}\right)$ 
\cite{Benenti-Carlo-Prosen} as 
\[
\left(\begin{array}{c}
q_{t+1}^{1}\\
p_{t+1}^{1}
\end{array}\right)=\mathcal{M}_{1}\left(\begin{array}{c}
q_{t}^{1}\\
p_{t}^{1}+\epsilon\left(q_{t}^{1}\right)+\kappa\left(q_{t}^{1},q_{t}^{2}\right)
\end{array}\right)
\]
and 
\[
\left(\begin{array}{c}
q_{t+1}^{2}\\
p_{t+1}^{2}
\end{array}\right)=\mathcal{M}_{2}\left(\begin{array}{c}
q_{t}^{2}\\
p_{t}^{2}+\epsilon\left(q_{t}^{2}\right)+\kappa\left(q_{t}^{1},q_{t}^{2}\right)
\end{array}\right)
\]
where the coupling between both maps is given by 
\[
\kappa\left(q_{t}^{1},q_{t}^{2}\right)=-\frac{K_{c}}{2\pi}\sin\left(2\pi q_{t}^{1}+2\pi q_{t}^{2}\right).
\]
The quantized version of the two degrees of freedom system is obtained 
as the tensor product of the quantized one degree of freedom maps, given 
by a $N^{2}\times N^{2}$ unitary matrix 
\[
U_{j_{1}j_{2},k_{1}k_{2}}^{2D}=U_{j_{1}k_{1}}U_{j_{2}k_{2}}C_{j_{1}j_{2}},
\]
with the coupling matrix (diagonal in the coordinate representation)
\[
C_{j_{1}j_{2}}=\exp\left\{ \left(\frac{iNK_{c}}{2\pi}\right)
\cos\left[\frac{2\pi}{N}\left(j_{1}+j_{2}\right)\right]\right\}, 
\]
where $j_{1},j_{2},k_{1},k_{2}\in\{0,\ldots,N-1\}$. We use $K=0.25$ and $K_c=0.5$ throughout 
our work, which guarantees the Anosov condition \cite{Ozorio 1994}.

\section{Results}

\label{sec4}

To investigate the quantum to classical correspondence regarding complexity growth, 
we study the evolution in time of $h[W]$ and of its classical counterpart $h[\rho_{c}]$. 
As initial states we have used a Gaussian 
phase space distribution with dispersion equal to $\sqrt{\hbar}$,
and its quantum analogue, a coherent state on the torus, for both degrees of freedom. Profiting from the fact 
that the latter is a pure state we just compute the von Neumann entropy which is half 
the WSE. In the following, when we refer to WSE and CSE we mean WSE/2 and CSE/2. 
We take $N=2^{6}$ for each map.

\begin{figure}[htp]
\hspace{0cm} \includegraphics[width=8cm]{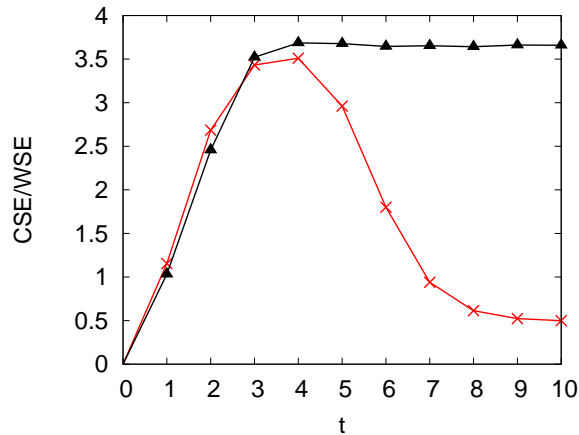} 
\caption{(Color online) CSE ((red) gray line with crosses) 
and WSE (black line with triangles) as a function of time t (in map steps) in 
the HH case for $N=2^{6}$. Initial distributions are centered at $(q,p)=(0.5,0.5)$.}
\label{fig1} 
\end{figure}

\begin{figure}[htp]
 \vspace{1cm}
 \includegraphics[width=8cm]{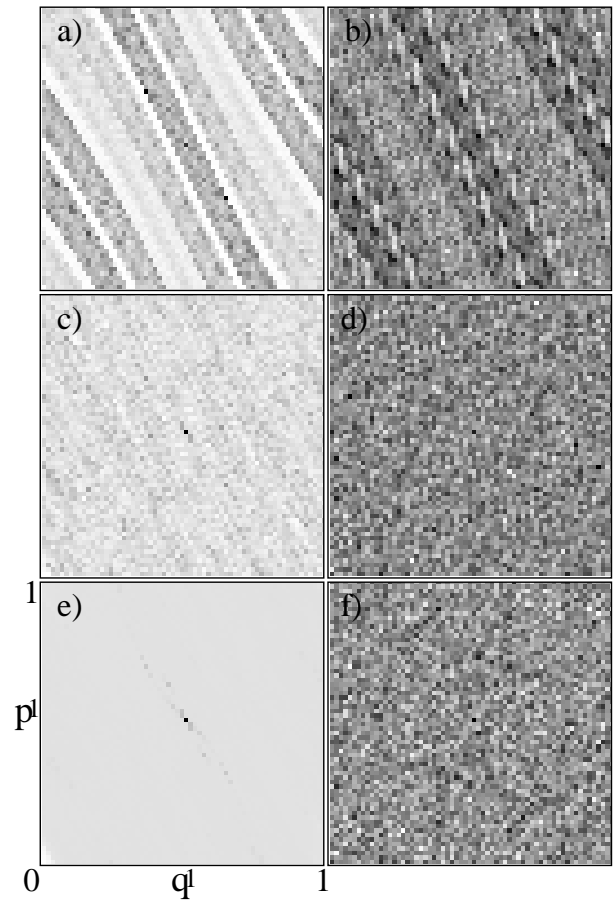} 
 \caption{Liouville and Wigner distributions at times $t=3$ (a) and b)), $t=4$ (c) and d)), and 
 $t=10$ (e) and f)), for the HH case with initial conditions centered at $(q,p)=(0.5,0.5)$.}
 \label{fig2} 
 \end{figure}

First, we consider the HH case with the initial distributions centered at $(q,p)=(0.5,0.5)$, 
which is a period 1 fixed point of both, the hyperbolic and the elliptic maps. 
CSE and WSE as a function of time (map steps) are displayed in Fig. \ref{fig1}. 
The Liouville and Wigner distributions develop similar structures of increasing complexity as 
the evolution takes place. At time $t=3$, and after growing at a rate given by the average Lyapunov exponent, 
they show the maximum complexity where 
features of the stable manifold are still visible in the quantum case (see Fig. \ref{fig2} a) and b)). 
From $t=3$ on the classical distribution becomes less 
complex due to discretization while the quantum one keeps its complexity through intertwined coherence patterns, 
as can be seen in Fig. \ref{fig2} c) and d). In case the classical distribution were not discretized, 
the CSE would continue to grow. The WSE grows until saturation 
at a value of the order of $\ln(0.6\times N)$, as predicted in \cite{lak2002}. At time $t=10$ the classical distribution 
is almost completely smoothed while the quantum one keeps the same morphology as the one at $t=4$ (see Fig. \ref{fig2} e) and f)) .
We notice that we have removed the effects of the torus periodicity on the Wigner distributions in all figures \cite{dittrich}.

\begin{figure}[htp]
\hspace{0cm} \includegraphics[width=8cm]{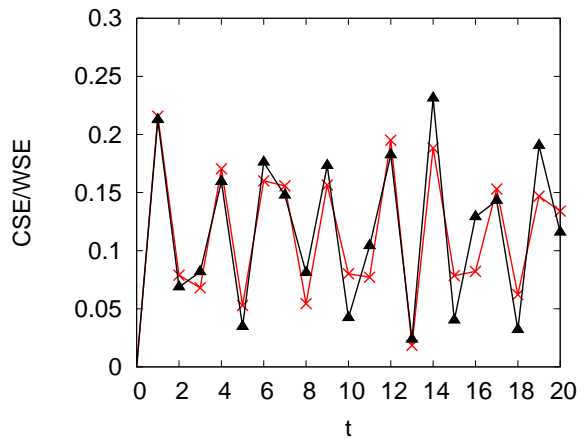} 
\caption{(Color online) 
CSE ((red) gray line with crosses) 
and WSE (black line with triangles) as a function of time t (in map steps) in 
the EE case for $N=2^{6}$. Initial distributions are centered at $(q,p)=(0.5,0.5)$.}
\label{fig3} 
\end{figure}

\begin{figure}[htp]
\hspace{0cm} \includegraphics[width=8cm]{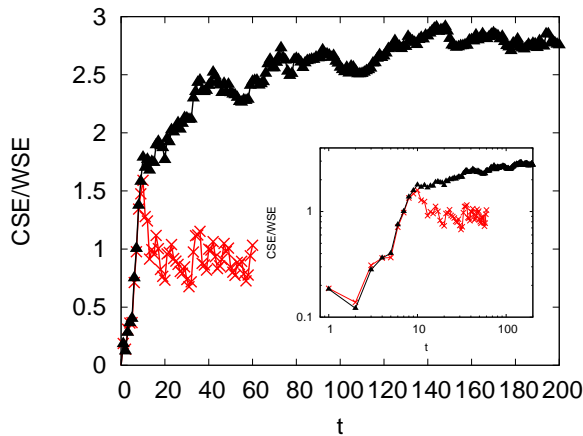} 
\caption{(Color online) 
CSE ((red) gray line with crosses) 
and WSE (black line with triangles) as a function of time t (in map steps) in 
the EE case for $N=2^{6}$. Initial distributions are centered at $(q,p)=(\pi/4,\pi/4)$.}
\label{fig4} 
\end{figure}

\begin{figure}[htp]
 \vspace{1cm}
 \includegraphics[width=8cm]{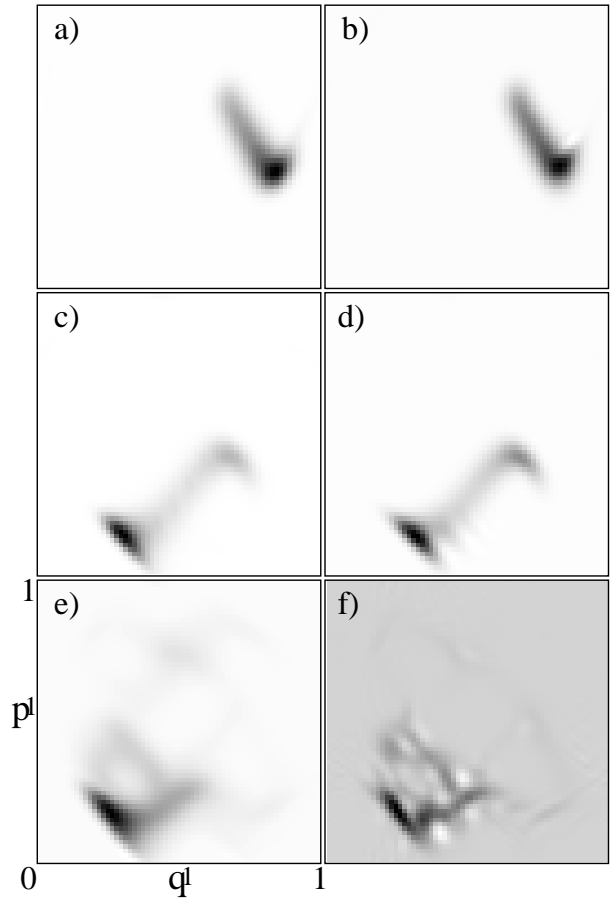} 
 \caption{Liouville and Wigner distributions at times $t=8$ (a) and b)), $t=11$ (c) and d)), and 
 $t=50$ (e) and f)), for the EE case with initial conditions centered at $(q,p)=(\pi/4,\pi/4)$.}
 \label{fig5} 
 \end{figure}

The EE case is highly dependent on where the initial conditions are taken. 
We first show the evolution of the CSE/WSE as a function of time for distributions 
centered at the previous values, i.e. at the fixed point of period 1. 
As can be seen in Fig. \ref{fig3} complexity does not grow significantly and the 
quantum and classical behaviour is remarkably similar at all times (from $t=3$ on the agreement 
worsens). Small oscillations reflect 
the rotation of the distributions which do not explore much of the phase space. 
But if we select initial distributions centered at $(q,p)=(\pi/4,\pi/4)$ for instance (we just take 
a representative case from the maximum complexity ones), complexity grows 
as shown in Fig. \ref{fig4}. The saturation values of the WSE are always lower than the one for the HH case, 
but greater than those of their corresponding CSE, suggesting that quantum effects begin to play an 
important role. In fact, from $t=10$ on, the two curves take completely different 
behaviours. Moreover, the WSE reaches its maximum value after approximately $160$ map steps, at a much slower 
rate than the HH case reflecting a different mechanism for complexity growth. In the inset we show the same quantities 
but in log-log scale. Here not only the power law behaviour becomes clear, but also the abrupt change of slope at $t=10$. 
If we look at Fig. \ref{fig5} we can see how the quantum distribution develops interference fringes 
from $t=8$ (panel b)) to $t=11$ (panel d)). The classical distribution just develops a secondary 
bulb of high density but of course coherences do not appear (see panels a) and c)). Finally, 
for $t=50$ we can already see the typical morphology of the quantum distribution that has developed 
a lot of fringes in contrast with the classical one (see panels e) and f)).
 It is interesting to note that up to now everything seems to agree with the results of 
\cite{nature2016} regarding the entangling power of chaotic and regular regions of the 
classical phase space, but keeping in mind this marked dependence on the initial conditions 
in the regular case. 

\begin{figure}[htp]
\hspace{0cm} \includegraphics[width=8cm]{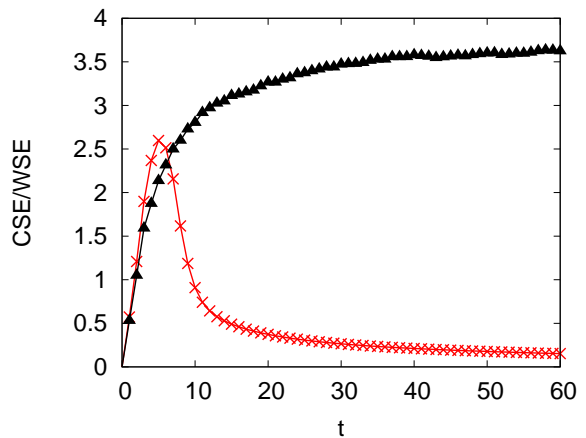} 
\caption{(Color online) 
CSE ((red) gray line with crosses) 
and WSE (black line with triangles) as a function of time t (in map steps) in 
the HE case for $N=2^{6}$. Initial distributions are centered at $(q,p)=(0.5,0.5)$.}
\label{fig6} 
\end{figure}

\begin{figure}[htp]
 \vspace{1cm}
 \includegraphics[width=8cm]{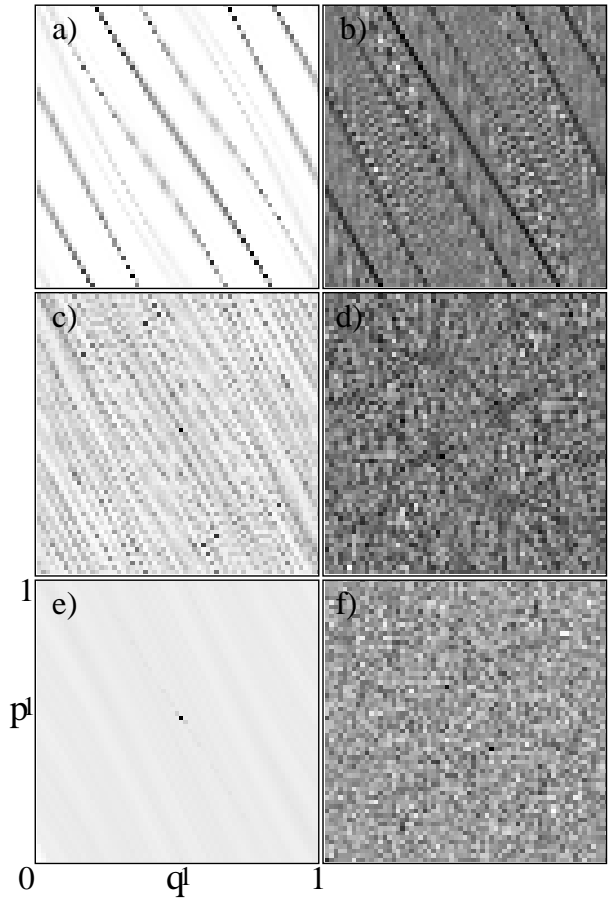} 
 \caption{Liouville and Wigner distributions at times $t=3$ (a) and b)), $t=5$ (c) and d)), and 
 $t=50$ (e) and f)), for the HE case with initial conditions centered at $(q,p)=(0.5,0.5)$.}
 \label{fig7} 
 \end{figure}

Finally, we analyze the HE case for which we take the initial distributions centered 
at $(q,p)=(0.5,0.5)$. The WSE saturates at values similar to the HH case, 
although it takes a much longer time to reach them (see Fig. \ref{fig6}). The CSE only grows until 
$t=5$ and then decreases due to discretization. 
It is remarkable that just one 
hyperbolic degree of freedom is enough to reach maximum complexity, although at a classical level 
the dynamic is not completely ergodic. In this sense the behaviour is strongly 
different than that of a mixed phase space with regular and chaotic regions. 
By looking at Fig. \ref{fig7} it becomes clear that the Liouville distribution at times $t=3$ and $t=5$ (see panels a) and c)) 
develop more and more complex structures associated to the stable manifold, but that at $t=50$ (see panel e)) it has 
already washed out almost all details. The corresponding quantum distributions (for the same times) 
at the right panels show a different mechanism of complexity growth mainly based on coherences after $t=3$.

\section{Conclusions}

\label{sec5}

We have studied a generic system consisting of two coupled 
perturbed cat maps, considering the doubly hyperbolic, elliptic, and the 
mixed cases. By using the WSE and the CSE as two sides of the same 
complexity notion we find that for the HH case, the 
quantum and classical complexity growth share the same behaviour 
(despite discretization effects of the classical distribution). The WSE and the CSE 
reach the maximum theoretical limit predicted in \cite{lak2002}, which is 
at the order of $\ln(0.6\times N)$. In the EE case this quantum to classical 
correspondence depends on the initial conditions, but the HH case 
complexity maximum is not reached. This confirms recent findings 
published in \cite{nature2016} regarding the production of entanglement
in chaotic and regular regions of the phase space. But it is important 
to clarify that entanglement entropy generation and classical chaotic or regular 
behaviour are directly related through the complexity notion, in these cases. 
Thus the connection is not surprising. 

Moreover, this is not always this way, as for instance in 
the non-generic baker map, which is chaotic but not complex \cite{Benenti-Carlo-Prosen}.
In the HE case of our generic system, the WSE reaches the maximum theoretical value 
around $\ln(0.6\times N)$, similarly to the HH 
case, although the time of the transient is much longer. The CSE does not 
reach this value and has a markedly different behaviour. This reveals that in the mixed 
scenario the quantum mechanisms of complexity growth (namely coherences) play a 
central role. On the other hand, just one hyperbolic degree of freedom is enough to reach
maximum complexity, despite the dynamics is not completely ergodic at the classical level.

Our results provide a wider picture of complexity growth, including mixed dynamics scenarios. 
These findings suggest new experiments with controllable quantum systems, where the 
behaviour of each component could be selected to be regular or chaotic. 
In the future, we will study the role played by complex
eigenvalues of the symplectic matrix leading to loxodromic behaviour \cite{multicat}.


\section*{Acknowledgments}


Support from CONICET is gratefully acknowledged.

\vspace{3pc}


\end{document}